\documentclass[preprint,12pt]{elsarticle}
\usepackage{lineno}
\usepackage{graphicx}
\usepackage{bm}
\usepackage{amsmath,upgreek}
\usepackage{hyperref}
\modulolinenumbers[5]

\journal{JMMM}

\begin{document}

\begin{frontmatter}

\title{Exactly solvable model for drift of suspended ferromagnetic particles induced by the Magnus force}
\author{S.I.~Denisov\corref{cor1}}
\ead{denisov@sumdu.edu.ua}
\cortext[cor1]{Corresponding author}
\author{B.O.~Pedchenko\corref{cor2}}
\author{O.V.~Kvasnina\corref{cor3}}
\author{E.S.~Denisova\corref{cor4}}
\address{Sumy State University, Rimsky-Korsakov Street 2, UA-40007
Sumy, Ukraine}

\begin{abstract}
The phenomenon of drift motion of single-domain ferromagnetic particles induced by the Magnus force in a viscous fluid is studied analytically. We use a minimal set of equations to describe the translational and rotational motions of these particles subjected to a harmonic force and a non-uniformly rotating magnetic field. Assuming that the azimuthal angle of the magnetic field is a periodic triangular function, we analytically solve the rotational equation of motion in the steady state and calculate the drift velocity of particles. We study in detail the dependence of this velocity on the model parameters, discuss the applicability of the drift phenomenon for separation of particles in suspensions, and verify numerically the analytical predictions.
\end{abstract}


\end{frontmatter}

\linenumbers

\section{Introduction}
\label{Intr}

The Magnus effect is a commonly observed effect, which usually refers to the lateral deviation of the trajectory of a spinning body moving through a medium from the trajectory of a non-spinning body. This effect plays an important role in the dynamics of spinning bodies, for example, in sport \cite{Meht2008, Cros2011}, aeronautics \cite{Seif2012} and planet formation \cite{YaKi2014, Forb2015}. It should be noted that the Magnus effect refers also to some localized objects (like vortices in magnets, superconductors and superfluids), but its origin is quite different from that for spinning bodies (see, e.g., Refs.~\cite{ThSt2006, Son2016} and references therein).

The lateral deviation of trajectories is usually associated with the so-called Magnus force acting on spinning particles. Since this force depends on many factors, such as the particle size, shape and roughness, characteristics of the translational and rotational motions, dynamics of the surrounding medium, etc., its calculation is not a trivial task. Moreover, under certain conditions (they can be realized, e.g., for a spinning sphere moving in a rarefied gas \cite{BSE2003, KCPY2014}) there may exist the inverse, rather than classical, Magnus effect, in which the Magnus force direction is opposite to that predicted by Bernoulli's principle. However, in the case of smooth spherical particles, whose translational and rotational motions are characterized by small Reynolds numbers, the Magnus effect is classical and the Magnus force is determined analytically \cite{RuKe1961}. Although in general these conditions are rather restrictive, they can be easily realized for small particles suspended in a viscous fluid.

This approach is especially useful for studying the translational and rotational dynamics of single-domain ferromagnetic particles in suspensions. One of the reasons is that these particles can be used in such biomedical applications as cell separation, drug delivery, and hyperthermia treatment \cite{PCJD2003, LFPR2008}. For hyperthermia, the rotational properties of particles and their magnetization play the most important role and are the subject of much current research. In particular, the rotational properties induced by linearly and circularly polarized (rotating) magnetic fields have already been extensively studied in Refs.~\cite{UsLi2012, UsUs2015} and \cite{LDRB2015, Usad2017}, respectively.

In contrast, the features of the translational dynamics of ferromagnetic particles subjected to a harmonic force and an oscillating magnetic field are not of primary interest. This is because, due to the smallness of the Magnus force, the particle displacement caused by this force during the field period is also small. However, if the external force and magnetic field are properly synchronized, then the Magnus force can induce directed transport (drift) of particles. This effect, which was predicted and numerically confirmed in Refs.~\cite{DPP2016, DePe2017}, is of special interest for particle separation since the magnitude and direction of the drift velocity can easily be controlled by external parameters. In this paper, we present a complete analytical solution of a minimal set of equations, which describes the translational and rotational motions of suspended ferromagnetic particles in a particular case, and provide a comprehensive analysis of the drift velocity.

\section{Minimal set of equations of motion}
\label{Eqs}

We consider a dilute suspension of single-domain ferromagnetic particles of spherical shape, each of which has the same radius $a$ and is characterized by the magnetization $\mathbf{M} = \mathbf{M} (t)$ ($M = |\mathbf{M}| = \mathrm{const}$). It is assumed that the magnetization is frozen along the particle easy axis (i.e., the anisotropy field is large enough) and the translational and rotational motions, which are induced by the external driving force $\mathbf{f}_{d} = \mathbf{f}_{d}(t)$ and magnetic field $\mathbf{H} = \mathbf{H}(t)$, occur in such a way that the corresponding Reynolds numbers are small. The last condition implies that the inertial effects in the particle dynamics can be neglected and the translational and rotational equations are reduced to the force and torque balance equations, $\mathbf{F} = 0$ and $\mathbf{T} = 0$, respectively. In the simplest case we have $\mathbf{F} = \mathbf{f}_{d} + \mathbf{f}_{f} + \mathbf{f}_{l}$ and $\mathbf{T} = \mathbf{t}_{m} + \mathbf{t} _{f}$, where $\mathbf{f} _{f}$ is the friction force, $\mathbf{f} _{l}$ is the Magnus lift force, $\mathbf{t}_{m}$ is the external mechanical torque, and $\mathbf{t}_{f}$ is the frictional torque. With the above assumptions, the friction force is determined by the Stokes law, $\mathbf{f}_{f} = -6\pi\eta a\mathbf{v}$ ($\eta$ is the dynamic viscosity of the host fluid, $\mathbf{v} = \mathbf{v} (t)$ is the linear particle velocity), and the external mechanical torque equals the magnetic field torque, i.e., $\mathbf{t}_{m} = (4\pi a^{3}/3) \mathbf{M} \times \mathbf{H}$ (the sign $\times$ denotes the vector product). Finally, according to \cite{RuKe1961}, the Magnus force and frictional torque acting on the rotating particle are given by the formulas $\mathbf{f}_{l} = \pi\rho a^{3} \boldsymbol {\omega} \times \mathbf{v}$ and $\mathbf{t}_{f} = -8\pi \eta a^{3} \boldsymbol{\omega}$, where $\rho$ is the fluid density and $\boldsymbol{\omega}$ is the angular particle velocity.

Equations $\mathbf{F} = 0$, $\mathbf{T} = 0$ and the kinematic differential equation $d\mathbf{M}/dt = \boldsymbol{ \omega} \times \mathbf{M}$ form a complete set of equations that determines $\mathbf{v}$, $\boldsymbol{\omega}$ and $\mathbf{M}$ as functions of time. Substituting the solution $\boldsymbol{ \omega} = (1/6\eta)\, \mathbf{M} \times \mathbf{H}$ of equation $\mathbf{T} = 0$ into other equations and assuming that $\mathbf{f}_{d} = f_{m} \sin{(\Omega t)}\, \mathbf{e} _{x}$ ($f_{m}$ and $\Omega$ are the amplitude and angular frequency of the driving force, $\mathbf{e} _{x}$ is the unit vector along the coordinate axis $x$), one obtains
\begin{subequations}\label{u,m}
\begin{gather}
    \mathbf{u} +\gamma \mathbf{u}\times
    (\mathbf{m} \times \mathbf{h}) =
    \sin{(2\pi \tau)}\, \mathbf{e}_{x},
    \label{u2}
    \\[3pt]
    \dot{\mathbf{m}} = -\alpha \mathbf{m}
    \times (\mathbf{m} \times \mathbf{h}).
    \label{m2}
\end{gather}
\end{subequations}
Here, $\mathbf{u} = \mathbf{v}/v_{m}$, $\mathbf{m} = \mathbf{M} /M$, and $\mathbf{h} = \mathbf{H}/H_{m}$ are the dimensionless particle velocity, magnetization, and magnetic field, respectively, $v_{m} = f_{m} /6\pi a\eta$, $H_{m} = \max{|\mathbf{H}|}$, the overdot denotes the derivative with respect to the dimensionless time $\tau = \Omega t/2\pi$, and the dimensionless parameters $\gamma$ and $\alpha$ are defined as
\begin{equation}
    \gamma = \frac{\rho a^{2}MH_{m}}{36
    \eta^{2}}, \qquad
    \alpha = \frac{\pi MH_{m}}{3 \eta \Omega}.
    \label{g,a}
\end{equation}
According to Eqs.~(\ref{u,m}), the parameter $\gamma$ characterizes the magnitude of the Magnus force, and the parameter $\alpha$ can be associated with the inverse rotational relaxation time.

For further simplification, we assume that both the reduced magnetic field $\mathbf{h}$ and the reduced magnetization $\mathbf{m}$ rotate non-uniformly in the $xy$ plane and are represented in the form
\begin{subequations}\label{h,m}
\begin{align}
    \mathbf{h} &= \cos{\psi}\,\mathbf{e}_{x} +
    \sin{\psi}\,\mathbf{e}_{y},
    \label{h}
    \\[3pt]
    \mathbf{m} &= \cos{\varphi}\,\mathbf{e}_{x} +
    \sin{\varphi}\,\mathbf{e}_{y},
    \label{m}
\end{align}
\end{subequations}
where the azimuthal angle of the magnetic field, $\psi = \psi(\tau + \phi/2\pi)$, is a given periodic function of $\tau$ satisfying the condition $\psi|_{1/2+ \tau} = -\psi|_{\tau}$ (and thus $\psi|_{1+ \tau} = \psi|_{\tau}$), $\phi \in [0, 2\pi]$ is the initial phase, and $\varphi = \varphi (\tau)$ is the azimuthal angle of the magnetization. With these representations, we find $\mathbf{m} \times \mathbf{h} = \sin{\chi}\, \mathbf{e} _{z}$ ($\chi = \psi - \varphi$ is the lag angle), $\mathbf{m} \times (\mathbf{m} \times \mathbf{h}) = \sin{\chi}\, (\sin{\varphi} \,\mathbf{e}_{x} - \cos{\varphi}\, \mathbf{e}_{y})$, and $\dot{\mathbf{m}} = -\dot{ \varphi}\, (\sin{\varphi}\, \mathbf{e} _{x} - \cos{\varphi}\, \mathbf{e} _{y})$. Therefore, from Eqs.~(\ref{u,m}) with $\gamma \ll 1$ (since $\gamma \sim a^{2}$, this condition is not too restrictive for suspended particles) we arrive at the desired set of equations
\begin{subequations}\label{u,chi}
\begin{gather}
    \mathbf{u} = (\mathbf{e}_{x} + \gamma
    \sin{\chi}\,\mathbf{e}_{y})\sin{(2\pi \tau)},
    \label{u3}
    \\[3pt]
    \dot{\chi} + \alpha\sin{\chi} =\dot{\psi}
    \label{eq_chi}
\end{gather}
\end{subequations}
(for definiteness, we choose $\chi|_{\tau = 0} =0$). This set of equations is minimal in the sense that it is the simplest one that describes the translational motion of non-uniformly rotating particles in a viscous fluid. A more complete analysis of conditions under which Eqs.~(\ref{u,chi}) hold, including those responsible for the deterministic approach, is presented in Ref.~\cite{DePe2017}.

Thus, to find the characteristics of the translational motion of suspended ferromagnetic particles, which is described by the particle velocity (\ref{u3}), we first need to solve Eq.~(\ref{eq_chi}) governing the evolution of the periodically driven overdamped pendulum. Some properties of its solutions have already been studied, e.g., in the context of Josephson junctions \cite{KSC1998, IsAl2010}. However, to the best of our knowledge, there are no known steady-state solutions of this equation. Since the drift velocity is defined in the steady state (see below), the search for exact steady-state solutions of Eq.~(\ref{eq_chi}) is of great importance in the analytical study of the transport properties of suspended particles.

\section{Particle trajectory and drift velocity}
\label{Drift}

According to (\ref{u3}), the particle position $\mathbf{S}(\tau) = \int_{0}^{\tau} \mathbf{u}(\tau') d\tau'$ at time $\tau$ can be written in the form
\begin{eqnarray}
    \mathbf{S}(\tau) \!\!\!&=&\!\!\! \frac{1}
    {2\pi}[1 - \cos{(2\pi\tau)}]\,\mathbf{e}_{x}
    \nonumber \\[3pt]
    &&\!\!\! +\, \gamma \int_{0}^{\tau}\sin{[
    \chi(\tau' + \phi/2\pi)]}\sin{(2\pi\tau')}
    d\tau' \mathbf{e}_{y}.
    \label{S}
\end{eqnarray}
Representing the dimensionless time $\tau$ as $\tau = n + \xi$ with $n=0,1,2,...$ and $\xi \in [0,1]$, let us first introduce the relative particle position by the formula $\mathbf{R}_{n} (\xi) = \mathbf{S}(n + \xi) - \mathbf{S}(n)$. Then, defining the particle trajectory in the steady state as $\mathbf{R}(\xi) = \lim_{n \to \infty} \mathbf{R}_{n} (\xi)$, one obtains
\begin{eqnarray}
    \mathbf{R}(\xi) \!\!\!&=&\!\!\! \frac{1}
    {2\pi}[1 - \cos{(2\pi\xi)}]\,\mathbf{e}_{x}
    \nonumber \\[3pt]
    &&\!\!\! +\, \gamma \int_{\phi/2\pi}^{\xi +
    \phi/2\pi}\sin{[
    \chi_{\mathrm{st}}(\xi')]}\sin{(2\pi\xi' -
    \phi)} d\xi' \mathbf{e}_{y},
    \label{R}
\end{eqnarray}
where $\chi_{\mathrm{st}} (\xi) = \lim_{n \to \infty} \chi(n + \xi)$ is the steady-state solution of Eq.~(\ref{eq_chi}) at $\phi=0$.

The dimensionless particle displacement during one period of the external force (recall, in our model the magnetic field has the same period) is given by $\mathbf{s} = \mathbf{R}(1)$. Using (\ref{R}), we get $s_{x}=0$ and, since $\chi_{\mathrm{st}} (0) = \chi_{\mathrm{st}} (1)$ and $\sin{[\chi_{ \mathrm{st}} (1/2 + \xi)]} = -\sin{[ \chi_{\mathrm{st}} (\xi)]}$,
\begin{equation}
    s_{y} = 2\gamma \int_{0}^{1/2} \sin{[\chi_{
    \mathrm{st}}}(\xi)] \sin{(2\pi \xi - \phi)}
    d\xi.
    \label{s_y}
\end{equation}
The last result shows that the joint action of the harmonic force and non-uniformly rotating magnetic field induces the drift of suspended particles along the axis $y$. Because of its definition, the displacement $s_{y}$ can be considered also as the dimensionless drift velocity of particles (the dimensional drift velocity is written as $v_{\mathrm{dr}} = v_{m}s_{y}$) \cite{DePe2017}. Our aim here is to find the exact analytical solution of Eq.~(\ref{eq_chi}) in the steady state and study the dependence of $s_{y}$ on the model parameters.

\section{Exact analytical results}
\label{Exact}

\begin{figure}[ht]
    \centering
    \includegraphics[width=\columnwidth]{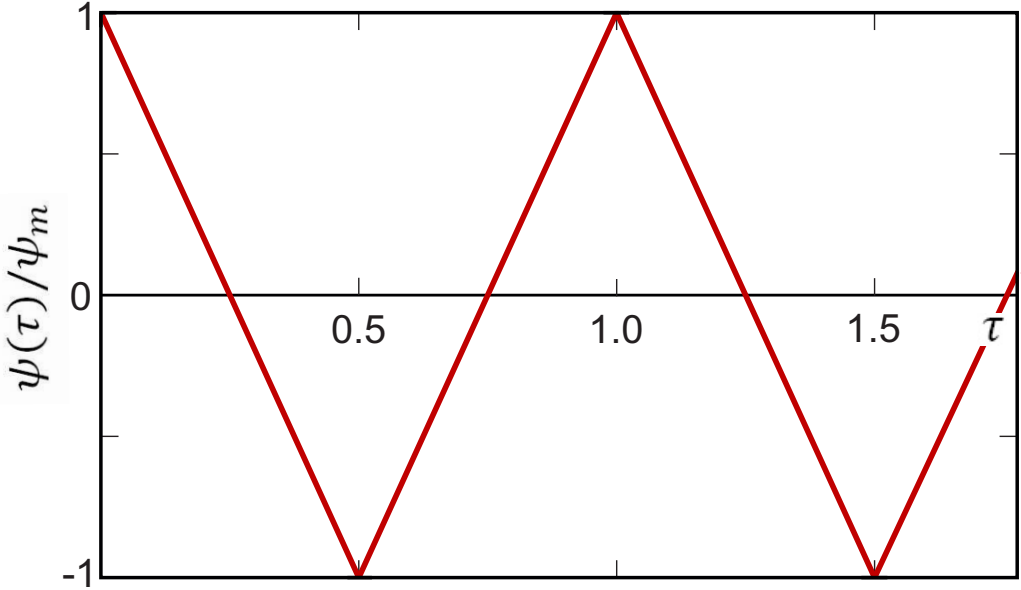}
    \caption{Plot of the function $\psi(\tau)
    /\psi_{m}$ defined in (\ref{def_psi}). This
    function is periodic with period 1 and
    satisfies the antisymmetry condition $\psi
    (1/2 + \tau) = - \psi(\tau)$.}
    \label{fig1}
\end{figure}

Next, we consider the case when the azimuthal angle $\psi(\tau)$ of the magnetic field is given by the periodic triangular function
\begin{equation}
    \psi(\tau) = \frac{2}{\pi}\psi_{m} \arcsin
    \left[ \cos {(2\pi \tau)}\right]
    \label{def_psi}
\end{equation}
($\psi_{m}>0$, see Fig.~\ref{fig1}). The main advantage of this choice of $\psi(\tau)$ is that the continuous steady-state solution of Eq.~(\ref{eq_chi}) can be found analytically. Indeed, according to (\ref{def_psi}), $\dot{\psi} (\tau)$ is the square wave, i.e., $\dot{\psi} (\tau) = \mp 4\psi_{m}$, where the upper sign corresponds to $\tau \in [n,n+1/2)$ ($n=0,1,2,...$) and the lower sign to $\tau \in [n+1/2, n+1)$. Therefore, in this case Eq.~(\ref{eq_chi}) (recall, we take $\phi=0$) can be rewritten in the differential form
\begin{equation}
    \frac{d\chi}{\sin{\chi} \pm \kappa} = -\alpha d\tau.
    \label{diff_chi}
\end{equation}
Here, $\kappa = 4\psi_{m}/\alpha$ is the dimensionless parameter characterizing the amplitude of the square wave. Because of the definition (\ref{s_y}), the steady-state solution of Eq.~(\ref{diff_chi}), $\chi_{ \mathrm {st}} (\xi) = \lim_{n \to \infty} \chi(n + \xi)$, is the most interesting for us. Taking into account that the function $\chi_{ \mathrm {st}} (\xi)$ is periodic with period 1, in the following we restrict the dimensionless time $\xi$ to the interval $[0,1]$. Then from Eq.~(\ref{diff_chi}) one immediately finds the following equations for $\chi_{ \mathrm {st}} (\xi)$:
\begin{subequations}\label{I1,I2}
\begin{align}
    \int_{\chi_{ \mathrm {st}} (0)}^{\chi_{
    \mathrm {st}} (\xi)}\frac{dx}{\sin{x}
    + \kappa} &= -\alpha \xi \quad \mathrm{as}
    \;\, \xi \in [0,1/2),
    \label{I1}
    \\[4pt]
    \int_{\chi_{ \mathrm {st}} (1/2)}^{\chi_{
    \mathrm {st}} (\xi)}\frac{dx}{\sin{x}
    - \kappa} &= -\alpha \left( \xi - \frac{1}{2}
    \right) \quad \mathrm{as}\;\, \xi \in [1/2,1).
    \label{I2}
\end{align}
\end{subequations}

Using the fact that $\dot{\psi}(1/2+ \xi) = -\dot{\psi}(\xi)$, it is easy to show from Eq.~(\ref{eq_chi}) that the condition
\begin{equation}
    \chi_{ \mathrm {st}} (1/2+\xi) = -2\pi l -
    \chi_{ \mathrm {st}} (\xi)
    \label{sym_cond}
\end{equation}
must hold, where $l$ is an integer number whose value is determined by the parameters $\alpha$ and $\kappa$ (see below). This means that Eq.~(\ref{I2}) can be excluded from further consideration and, to find $\chi_{ \mathrm {st}} (\xi)$ on the interval $[0,1]$, we can use Eq.~(\ref{I1}) and the condition (\ref{sym_cond}). Since the integral in Eq.~(\ref{I1}) depends on whether the parameter $\kappa$ is less, equal or greater than 1, we consider the drift motion of particles for these cases separately.

\subsection{Drift motion at \texorpdfstring{$\kappa<1$}{k}}
\label{kappa<1}

A simple analysis of Eqs.~(\ref{I1,I2}) shows that in this case the steady-state solution of Eq.~(\ref{diff_chi}) satisfies the following inequalities: $-\arcsin{ \kappa}\! < \chi_{ \mathrm {st}}(\xi) < \arcsin{\kappa}$. From this it follows that $l=0$, and so the condition (\ref{sym_cond}) at $\kappa <1$ reduces to $\chi_{ \mathrm {st}} (1/2+\xi) = - \chi_{ \mathrm {st}} (\xi)$. In order to find $\chi_{ \mathrm {st}} (\xi)$ for $\xi \in [0,1/2)$, we use Eq.~(\ref{I1}) and the standard integral
\begin{equation}
    \int \frac{dx}{\sin{x} + \kappa} = \frac{1}
    {\sqrt{1 - \kappa^{2}}} \ln \frac{\kappa
    \tan{(x/2)} + 1 - \sqrt{1 - \kappa^{2}}}{
    \kappa \tan{(x/2)} + 1 + \sqrt{1 -
    \kappa^{2}}}
    \label{int1}
\end{equation}
(see Eq.~(1.5.9.14) in Ref.~\cite{PBM1986}). Introducing the notations
\begin{equation}
    Q(\xi) = \kappa \tan{\frac{\chi_{\mathrm{st}}
    (\xi)}{2}} + 1
    \label{defQ}
\end{equation}
and
\begin{equation}
    \sigma = \frac{\alpha}{4} \sqrt{1 - \kappa^{2}}
    \label{sigma}
\end{equation}
and using (\ref{int1}), Eq.~(\ref{I1}) can be represented in the form
\begin{equation}
    \frac{Q(\xi) - \sqrt{1-\kappa^{2}}}{Q(\xi) +
    \sqrt{1-\kappa^{2}}} = \frac{Q(0) - \sqrt{1-
    \kappa^{2}}}{Q(0) + \sqrt{1-\kappa^{2}}}
    \,e^{-4\sigma\xi}.
    \label{eqQ}
\end{equation}
The solution of this equation with respect to $Q(\xi)$ yields
\begin{equation}
    Q(\xi) = \sqrt{1- \kappa^{2}}\, \frac{Q(0) +
    \sqrt{1- \kappa^{2}} \tanh{(2\sigma\xi)}}{Q(0)
    \tanh{(2\sigma\xi)} + \sqrt{1- \kappa^{2}}},
    \label{Q1}
\end{equation}
where $Q(0)$ can also be determined from Eq.~(\ref{eqQ}). Indeed, putting in this equation $\xi=1/2$ and replacing $Q(1/2)$ by $2-Q(0)$ [this is possible because $\chi_{ \mathrm {st}} (1/2) = -\chi_{ \mathrm {st}} (0)$], one gets
\begin{equation}
    Q(0) = \kappa^{2} \frac{\tanh{\sigma}}
    {\sqrt{1 - \kappa^{2}\big/ \cosh^{2}
    {\sigma}} + \sqrt{1- \kappa^{2}}} + 1.
    \label{Q(0)1}
\end{equation}
Now, using the notation (\ref{defQ}), the lag angle in the steady state can be represented as follows:
\begin{equation}
    \chi_{\mathrm{st}}(\xi) = 2\arctan{
    \frac{Q(\xi) -1} {\kappa}},
    \label{chi1}
\end{equation}
where $\arctan{x}$ denotes the principal value of the inverse tangent, i.e., $\arctan{x}$ belongs to the interval $(-\pi/2, \pi/2)$. Note also that in the particular case when $\xi=0$, expressions (\ref{chi1}) and (\ref{Q(0)1}) yield the result
\begin{equation}
    \chi_{\mathrm{st}}(0) = \arctan\left(
    \frac{\kappa} {\sqrt{1- \kappa^{2}}}
    \tanh{\sigma} \right).
    \label{chi(0)1}
\end{equation}
It shows that the condition $\chi_{\mathrm{st}}(0) \in [0, \pi/2)$ always holds.

\begin{figure}[ht]
    \centering
    \includegraphics[width=\columnwidth]{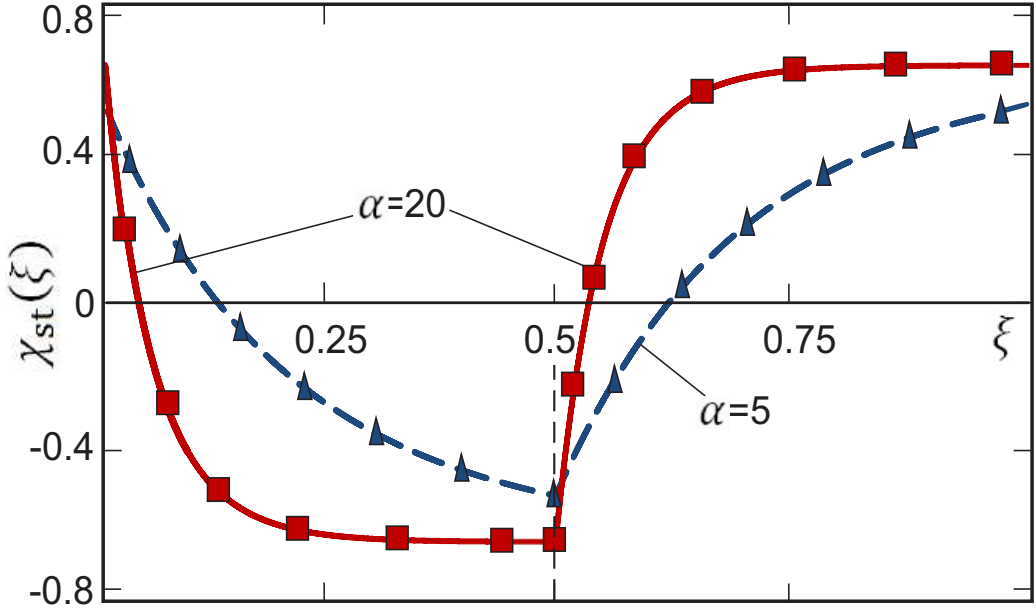}
    \caption{Time dependencies of the lag angle in
    the steady state at $\kappa = 0.6$. The solid
    and dashed lines represent the theoretical result
    (\ref{chi1}) for $\alpha = 20$ and $\alpha =5$,
    respectively. For the same values of $\alpha$,
    the numerical solutions of Eq.~(\ref{eq_chi}) on
    the $n$-th period of the function (\ref{def_psi})
    (to reach the steady state, we chose $n = 10^{2}$)
    are represented by square and triangle symbols.}
    \label{fig2}
\end{figure}

\begin{figure}[ht]
    \centering
    \includegraphics[width=\columnwidth]{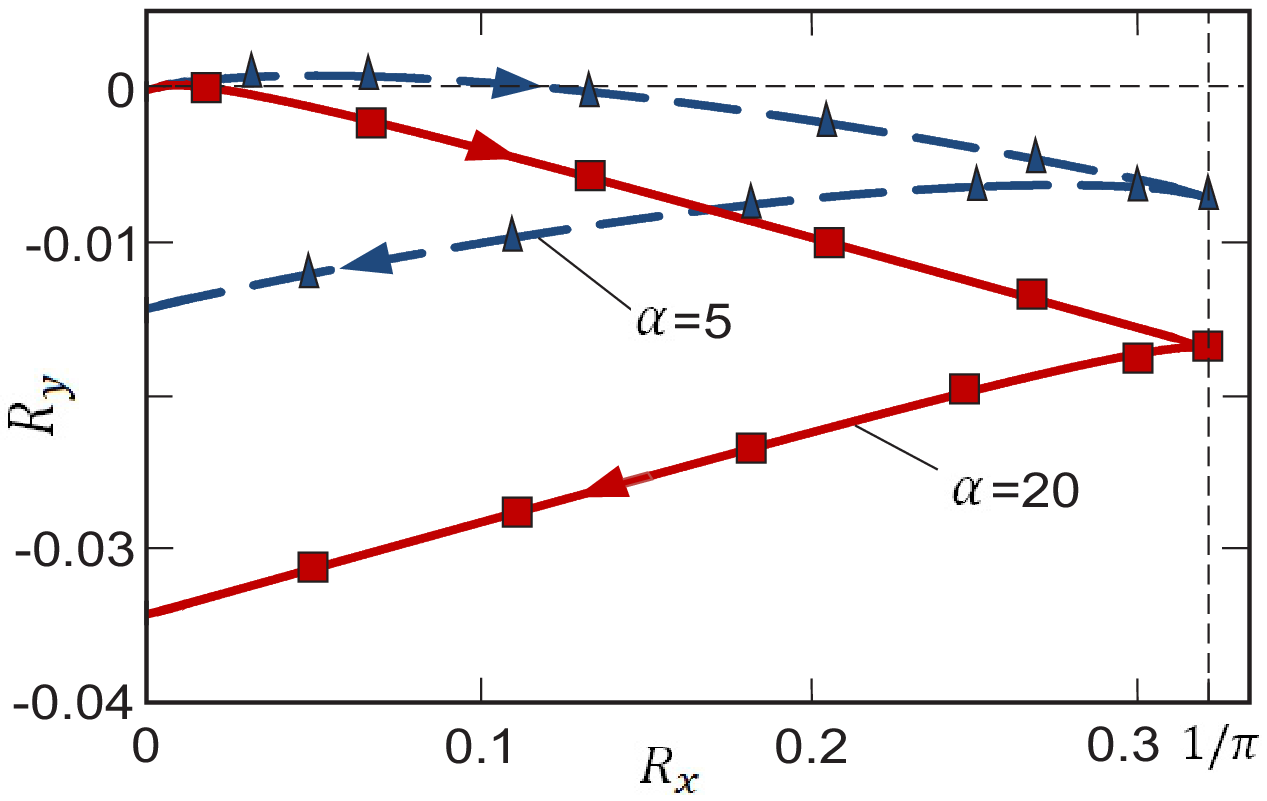}
    \caption{Fragments of the steady-state particle
    trajectories that are derived at $\kappa = 0.6$,
    $\gamma = 0.1$ and $\phi =0$ for one period of
    the driving force. The directed solid and dashed
    lines represent the theoretical trajectories obtained
    from (\ref{R}) and (\ref{sin(chi)1}) for $\alpha
    =20$ and $\alpha =5$. The initial point of these
    trajectories corresponds to $\xi =0$ ($R_{x}|_{\xi=0}
    = R_{y}|_{\xi=0} =0$) and the end points correspond
    to $\xi=1$ and determine the drift velocity ($R_{x}
    |_{\xi=1} =0$, $R_{y}|_{\xi=1} = s_{y}$). The
    square and triangle symbols represent the particle
    positions calculated numerically from (\ref{R})
    and Eq.~(\ref{eq_chi}).}
    \label{fig3}
\end{figure}

The theoretical plots of $ \chi_{\mathrm{st}}(\xi)$ on the interval $[0,1]$, which are obtained from (\ref{Q1})--(\ref{chi1}) and the condition $\chi_{ \mathrm {st}} (1/2+\xi) = - \chi_{ \mathrm {st}} (\xi)$, are shown in Fig.~\ref{fig2} by solid and dashed lines. The numerical results, which are represented in this figure by square and triangle symbols, confirm the theoretical ones. The influence of the Magnus force on the trajectories of magnetic particles, leading to their drift, is illustrated in Fig.~\ref{fig3}. As before, our theoretical and numerical results are in complete agreement.

Finally, let us calculate the drift velocity of ferromagnetic particles induced by the Magnus force. Taking into account that $\sin{(2 \arctan{x})} = 2x/(1+x^{2})$, from (\ref{chi1}) one obtains
\begin{equation}
    \sin\left[\chi_{\mathrm{st}}(\xi)\right] =
    2\kappa \frac{Q(\xi) -1}{\kappa^{2} +
    (Q(\xi) -1)^{2}}.
    \label{sin(chi)1}
\end{equation}
Substituting the right-hand side of this relation into (\ref{s_y}), we arrive at the following expression for the drift velocity:
\begin{equation}
    s_{y} = 4\gamma \kappa \int_{0}^{1/2}
    \frac{Q(\xi) -1}{\kappa^{2} + (Q(\xi)
    -1)^{2}} \sin{(2\pi \xi - \phi)}d\xi,
    \label{s_y1}
\end{equation}
which holds for $\kappa<1$. In the limiting case when $\alpha \to \infty$, we obtain in accordance with expressions (\ref{Q1}) and (\ref{s_y1}) that $\left.Q(\xi) \right|_{\alpha = \infty} = \sqrt{1 - \kappa^{2}}$ ($\xi \neq 0$) and
\begin{equation}
    \left.s_{y} \right|_{\alpha = \infty} = -
    \frac{2}{\pi}\, \gamma \kappa \cos{\phi}.
    \label{lim1}
\end{equation}
In the opposite limiting case, when $\alpha \to 0$, we have $Q(\xi) \sim 1 + (\kappa^{2}/8)(1 - 4\xi) \alpha$ and
\begin{equation}
    \left.s_{y} \right|_{\alpha \to 0} \sim -
    \frac{1} {\pi^{2}}\, \gamma \kappa\alpha
    \sin{\phi}.
    \label{lim2}
\end{equation}

\begin{figure}[ht]
    \centering
    \includegraphics[width=\columnwidth]{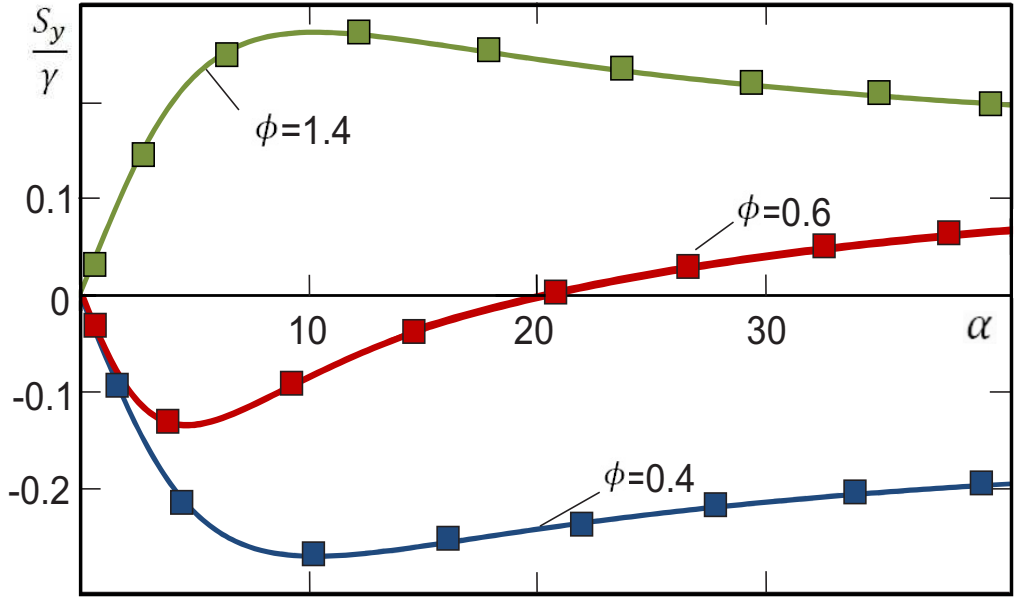}
    \caption{Reduced drift velocity $s_{y}/\gamma$ as
    a function of the parameter $\alpha$. The solid lines,
    which illustrate the dependence of this function on the
    initial phase $\phi$ (measured in radians), represent
    the theoretical result (\ref{s_y1}) at $\kappa = 0.6$.
    The numerical results derived from (\ref{s_y}) and
    Eq.~(\ref{eq_chi}) are shown by square symbols.}
    \label{fig4}
\end{figure}

\begin{figure}[ht]
    \centering
    \includegraphics[width=\columnwidth]{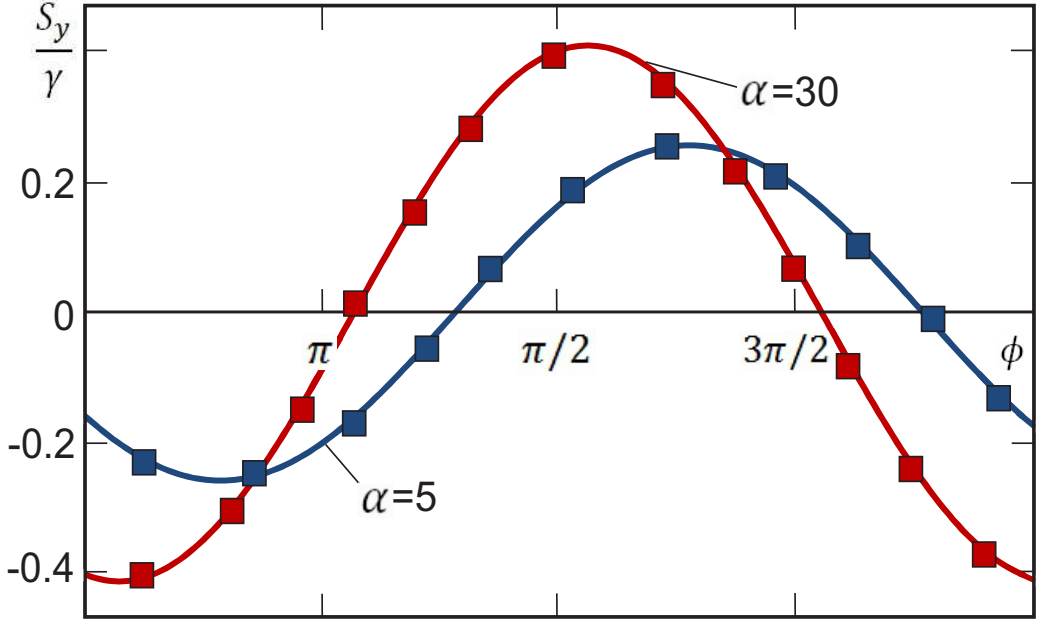}
    \caption{Reduced drift velocity $s_{y}/\gamma$ as
    a function of the initial phase $\phi$ for $\kappa
    = 0.6$ and different values of $\alpha$. The
    theoretical and numerical results are determined
    in the same way as in Fig.~\ref{fig4}.}
    \label{fig5}
\end{figure}

The dependencies of the drift velocity (\ref{s_y1}) on the parameter $\alpha$ for different values of the initial phase $\phi$ are illustrated in Fig.~\ref{fig4}. Their remarkable feature is the existence of a local maximum or minimum. Because $\alpha$ depends on the angular frequency $\Omega$, see (\ref{g,a}), this means that there is such a frequency at which the absolute value of the drift velocity reaches a maximum. Another notable feature is that the drift of different particles (i.e., particles characterized by different values of the parameter $\alpha$) may occur in opposite directions, if the initial phase is chosen appropriately. Specifically, the reference situation is represented in Fig.~\ref{fig4} by the theoretical and numerical data obtained for $\phi = 0.6\, \mathrm{rad}$. In this case, all particles with $\alpha < \alpha_{\mathrm{cr}}$, where $\alpha_{ \mathrm{cr}} \approx 20.58$ is the solution of equation $s_{y}=0$, drift against the $y$ axis ($s_{y}<0$), and all particles with $\alpha > \alpha_{ \mathrm{cr}}$ drift along this axis ($s_{y}>0$). Recently, we proposed to use this phenomenon for separation of core-shell ferromagnetic particles in dilute suspensions \cite{DePe2017}. It should also be emphasized that, according to (\ref{s_y1}), the initial phase of the magnetic field strongly influences the drift velocity. As seen in Fig.~\ref{fig5}, the initial phase controls both its magnitude and direction.

\subsection{Drift motion at \texorpdfstring{$\kappa=1$}{k}}
\label{kappa=1}

In order to find the drift velocity at $\kappa =1$, we first take the limit $\kappa \to 1$ in the above expressions. Since $\tanh{x} \sim x$ and $\cosh{x} \sim 1$ as $x \to 0$, from (\ref{Q1}) and (\ref{Q(0)1}) one gets
\begin{equation}
    \left.Q(\xi)\right|_{\kappa = 1} =
    \frac{2\left.Q(0)\right|_{\kappa = 1}}
    {\alpha \left.Q(0)\right|_{\kappa = 1}\xi + 2}
    \label{Q2}
\end{equation}
and
\begin{equation}
    \left.Q(0)\right|_{\kappa = 1} = \frac{1 +
    \alpha/4 + \sqrt{1 + (\alpha/4)^{2}}}{1 +
    \sqrt{1 + (\alpha/4)^{2}}}
    \label{Q(0)2}
\end{equation}
[in addition, formula (\ref{chi(0)1}) yields $\left.\chi_{ \mathrm{st}}(0) \right|_{\kappa = 1} = \arctan{( \alpha /4)}$]. Fur\-ther, using (\ref{Q2}) and (\ref{Q(0)2}), we can write from (\ref{sin(chi)1})
\begin{equation}
    \left.\sin\left[\chi_{\mathrm{st}}(\xi)\right]
    \right|_{\kappa = 1} = -1 + \frac{2}{1 + q^{2}
    (\xi)},
    \label{sin(chi)2}
\end{equation}
where
\begin{equation}
    q(\xi) = \alpha\xi - \frac{\alpha}{4}
    + \sqrt{1 + \left( \frac{\alpha}{4} \right)^{2}},
    \label{def_q}
\end{equation}
and, as a consequence, represent the particle drift velocity (\ref{s_y1}) at $\kappa =1$ in the form
\begin{equation}
    \left.s_{y}\right|_{\kappa = 1} = -\frac{2\gamma}{\pi}
    \cos{\phi} + 4\gamma \int_{0}^{1/2} \frac{
    \sin{(2\pi\xi -\phi)}}{1 + q^{2}(\xi)} d\xi.
    \label{s_y2}
\end{equation}
Note that the same result follows from the direct solution of Eq.~(\ref{eq_chi}) for $\kappa = 1$ \cite{DePe2017}.

\subsection{Drift motion at \texorpdfstring{$\kappa>1$}{k}}
\label{kappa>1}

Under this condition, the solution of Eq.~(\ref{I1}) can strongly differ from that obtained for $\kappa \leq 1$. The reason is that the integrand in (\ref{I1}) is restricted for all $x$ and, as a consequence, the lag angle $\chi_{\mathrm{st}} (\xi)$ can vary in a wide interval. In other words, if the parameters $\alpha$ and $\kappa$ are sufficiently large (see below), then $l$ in (\ref{sym_cond}) can be non-zero. To find $\chi_{\mathrm{st}} (\xi)$ in this case, we use the standard integral (see Eq.~(1.5.9.13) in Ref.~\cite{PBM1986})
\begin{equation}
    \int \frac{dx}{\sin{x} + \kappa} = \frac{2}
    {\sqrt{\kappa^{2} - 1}} \arctan \frac{\kappa
    \tan{(x/2)} + 1}{\sqrt{\kappa^{2} - 1}}
    \label{int2}
\end{equation}
($\kappa > 1$). With this result, Eq.~(\ref{I1}) can be reduced to
\begin{equation}
    \tan\left( \arctan \frac{R(0)}{\sqrt{
    \kappa^{2} - 1}} - \arctan \frac{R(\xi)}
    {\sqrt{ \kappa^{2} - 1}} \right) =
    \tan{(2\nu\xi)},
    \label{eqR1}
\end{equation}
where, by definition,
\begin{equation}
    R(\xi) = \kappa \tan{\frac{\chi_{\mathrm{st}}
    (\xi)}{2}} + 1
    \label{defR}
\end{equation}
(i.e., $R(\xi)$ is $Q(\xi)$ at $\kappa > 1$) and
\begin{equation}
    \nu = \frac{\alpha}{4} \sqrt{\kappa^{2}-1}.
    \label{nu}
\end{equation}
Solving equation
\begin{equation}
    \sqrt{\kappa^{2} - 1}\, \frac{R(0) -
    R(\xi)}{\kappa^{2} - 1 + R(0)R(\xi)} =
    \tan{(2\nu\xi)}
    \label{eqR2}
\end{equation}
with respect to $R(\xi)$, which follows from Eq.~(\ref{defQ}), one finds
\begin{equation}
    R(\xi) = \sqrt{\kappa^{2} -1}\, \frac{R(0) -
    \sqrt{\kappa^{2}- 1} \tan{(2\nu\xi)}}
    {R(0) \tan{(2\nu\xi)} + \sqrt{\kappa^{2}-
    1}}.
    \label{R1}
\end{equation}

The initial value of this function, $R(0)$, obeys the equation
\begin{equation}
    \frac{2\sqrt{\kappa^{2} - 1}\,[R(0) - 1]}
    {\kappa^{2} - 1 + R(0)[2 - R(0)]} =
    \tan{\nu},
    \label{eqR(0)}
\end{equation}
which can be derived from Eq.~(\ref{eqR2}) by putting $\xi=1/2$ and using condition (\ref{sym_cond}). Assuming that $\chi_{\mathrm{st} }(0) \in (0, \pi)$, the solution of Eq.~(\ref{eqR(0)}) that satisfies the condition $R(0)>1$ is given by
\begin{equation}
    R(0) = 1 - \frac{\sqrt{\kappa^{2} -1}}{\tan
    {\nu}} + \sqrt{ \frac{\kappa^{2} -1}
    {\tan^{2} {\nu}} + \kappa^{2}}.
    \label{R(0)}
\end{equation}
If the parameter $\nu$ is bounded by
\begin{equation}
    p\pi < \nu < (p+1)\pi
    \label{cond1}
\end{equation}
($p=0,1,2,...$), then from (\ref{defR}), (\ref{R(0)}) and the identity $\tan{x} = 2\tan{(x/2)}/[1- \tan^{2}{(x/2)}]$ we straightforwardly obtain
\begin{equation}
    \chi_{\mathrm{st}}(0) = \arctan{\left(
    \frac{\kappa}{\sqrt{\kappa^{2} - 1}}
    \tan{\nu}\right)}
    \label{chi(0)2}
\end{equation}
for $p\pi < \nu < p\pi + \pi/2$ and
\begin{equation}
    \chi_{\mathrm{st}}(0) = \pi + \arctan{
    \left( \frac{\kappa}{\sqrt{\kappa^{2} -
    1}} \tan{\nu}\right)}
    \label{chi(0)3}
\end{equation}
for $p\pi + \pi/2 < \nu < p\pi + \pi$. Note that, since according to (\ref{chi(0)2}) and (\ref{chi(0)3}) $\left. \chi_{\mathrm{st}} (0) \right|_{\nu = p\pi + \pi/2 -0} = \left. \chi_{\mathrm{st}}(0) \right|_{\nu = p\pi + \pi/2 +0}$, $\chi_{\mathrm{st}}(0)$ is a continuous function of the parameters $\alpha$ and $\kappa$.

As it follows from definition (\ref{defR}) and the fact that the lag angle in the steady state, $\chi_{\mathrm{st}}(\xi)$, monotonically decreases in the interval $[0,1/2)$, the function $\chi_{\mathrm{st}}(\xi)$ at the condition (\ref{cond1}) can be expressed through $R(\xi)$ in the following way:
\begin{equation}
    \chi_{\mathrm{st}}(\xi) = -2\pi p(\xi) +
    2\arctan \frac{R(\xi) -1}{\kappa}.
    \label{chi2}
\end{equation}
Here, $p(\xi)$ is an integer-valued function of the variable $\xi$ [recall, $\xi \in [0,1/2)$], which provides continuity of $\chi_{\mathrm {st}}(\xi)$. Specifically, if $0 < \nu < \pi$ (i.e., $p=0$), then $p(\xi) = 0$ for all permissible values of $\xi$, and if $p\geq 1$, then $p(\xi)$  is determined as
\begin{equation}
    p(\xi) = \left\{\!\! \begin{array}{ll}
    0, & 0 \leq \xi < \xi_{1},
    \\ [3pt]
    1, & \xi_{1} < \xi < \xi_{2},
    \\
    \vdots & \vdots
    \\
    p, & \xi_{p} < \xi < 1/2,
    \end{array}
    \right.
    \label{n(xi)}
\end{equation}
where
\begin{equation}
    \xi_{k} = \frac{1}{2\nu} \left( k\pi -
    \arctan{\frac{\sqrt{\kappa^{2} - 1}}
    {R(0)}}\right)
    \label{xi_k}
\end{equation}
with $k=1,2,...,p$ are solutions of the equation $\tan{(2\nu \xi)} = - \sqrt{ \kappa^{2} - 1}/R(0)$ that belong to the interval $[0, 1/2)$. It should be emphasized that, according to (\ref{R1}), $|R(\xi_{k})| = \infty$ and, since $p(1/2) = p$, the integer number $l$ in the condition $\chi_{ \mathrm{st}}(1/2) = - 2\pi l - \chi_{ \mathrm{st}}(0)$, which follows from (\ref{sym_cond}), is equal to $p$.

\begin{figure}[ht]
    \centering
    \includegraphics[width=\columnwidth]{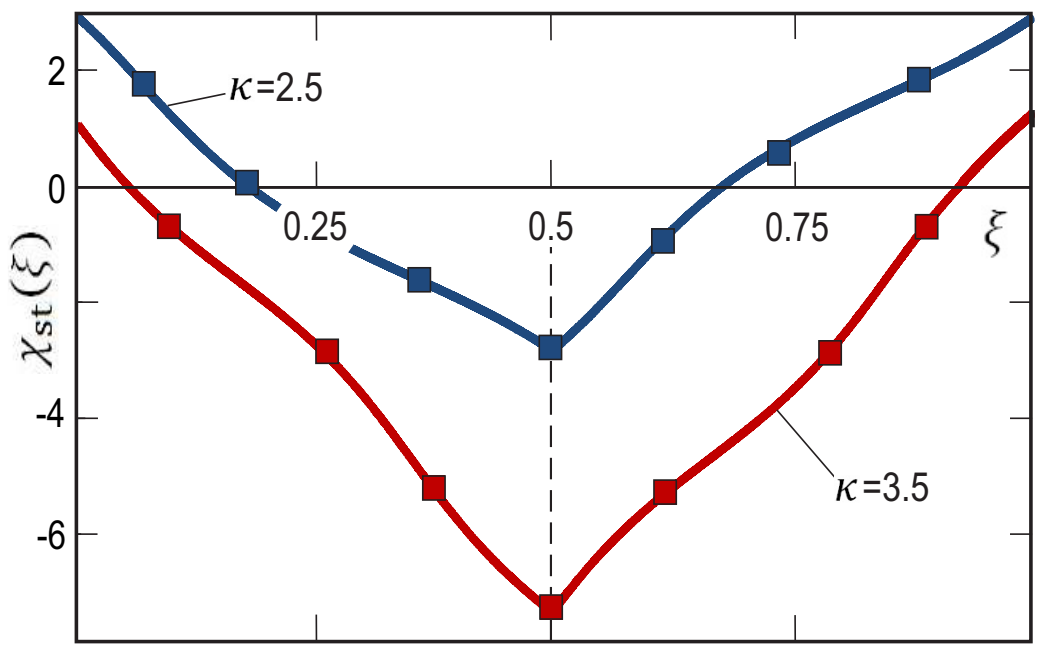}
    \caption{Lag angle in the steady state as a
    function of the dimensionless time $\xi$ for
    $\alpha = 5$, $\phi=0$ and different $\kappa$.
    The solid lines represent the theoretical result
    (\ref{chi2}), and the square symbols represent
    the numerical solution of Eq.~(\ref{eq_chi})
    satisfying the initial condition $\chi(0) = 0$.
    The plot for $\kappa = 2.5$ corresponds to $p=0$
    [because in this case $\nu \approx 2.86 \in (0,
    \pi)$], and the plot for $\kappa = 3.5$ corresponds
    to $p=1$ [because in this case $\nu \approx 4.19
    \in (\pi, 2\pi)$].}
    \label{fig6}
\end{figure}

For illustration, in Fig.~\ref{fig6} we show (solid lines) the dependence of the lag angle (\ref{chi2}) on the dimensionless time $\xi$ for $\alpha = 5$ and two values of the parameter $\kappa$. Namely, if $\kappa = 2.5$, i.e., $\nu \approx 2.86 \in (0, \pi)$, then the function $p(\xi)$ equals zero and, since $p=l=0$, in this case the lag angle satisfies the condition $\chi_{ \mathrm{st}}(1/2 + \xi) = - \chi_{ \mathrm{st}}(\xi)$. In contrast, if $\kappa = 3.5$, i.e., $\nu \approx 4.19 \in (\pi, 2\pi)$, then $p(\xi) = 0$ as $0 \leq \xi < \xi_{1}$ and $p(\xi) = 1$ as $\xi_{1} < \xi < 1/2$, where, according to (\ref{xi_k}), $\xi_{1} \approx 0.28$. Therefore, in this case $p=l=1$ and, as a consequence, the condition $\chi_{ \mathrm{st}}(1/2 + \xi) = -2\pi - \chi_{\mathrm{st}} (\xi)$ must hold. The numerical values for $\chi_{ \mathrm{st}}(\xi)$, obtained by numerical solution of Eq.~(\ref{eq_chi}) with the initial condition $\chi(0) = 0$, are represented by square symbols.

To calculate the drift velocity at $\kappa > 1$, we first express $\sin\left[ \chi_{ \mathrm{st}} (\xi)\right]$ in terms of $R(\xi)$. Proceeding as in Sec.~\ref{kappa<1}, we get
\begin{equation}
    \sin\left[\chi_{\mathrm{st}}(\xi)\right] =
    2\kappa \frac{R(\xi) -1}{\kappa^{2} +
    (R(\xi) -1)^{2}},
    \label{sin(chi)3}
\end{equation}
and hence the particle drift velocity (\ref{s_y}) takes the form
\begin{equation}
    s_{y} = 4\gamma \kappa \int_{0}^{1/2}
    \frac{R(\xi) -1}{\kappa^{2} + (R(\xi)
    -1)^{2}} \sin{(2\pi \xi - \phi)}d\xi.
    \label{s_y3}
\end{equation}
It is worthwhile to note that in the limits $\alpha \to 0$ and $\kappa \to 1$ this result reduces to (\ref{lim2}) and (\ref{s_y2}), respectively.

\begin{figure}[ht]
    \centering
    \includegraphics[width=\columnwidth]{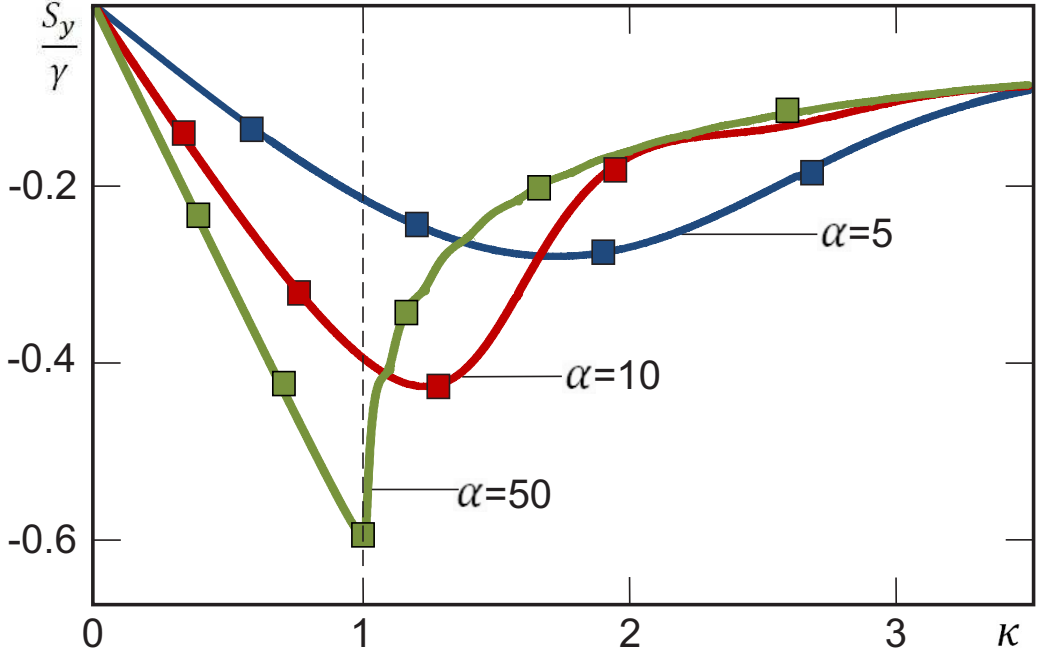}
    \caption{Reduced drift velocity $s_{y}/\gamma$
    as a function of the parameter $\kappa$ for
    $\phi = 0$ and different values of $\alpha$.
    The theoretical  plots at $\kappa <1$ and
    $\kappa >1$ are calculated using (\ref{s_y1})
    and (\ref{s_y3}), respectively, and the numerical
    results (square symbols) are obtained from the
    definition (\ref{s_y}) and Eq.~(\ref{eq_chi}).}
    \label{fig7}
\end{figure}

Our analysis showed that the dependencies of the drift velocity (\ref{s_y3}) on the parameter $\alpha$ and initial phase $\phi$ are qualitatively the same as for $\kappa < 1$. Therefore, we analyze here in more detail only the $\kappa$ dependence of $s_{y}$, which is illustrated in Fig.~\ref{fig7}. The theoretical results for $s_{y}$ (solid lines) are derived from (\ref{s_y1}) (when $\kappa <1$) and (\ref{s_y3}) (when $\kappa >1$), and the numerical ones (square symbols) are obtained from (\ref{s_y}) by numerical solution of Eq.~(\ref{eq_chi}). The most important feature of $s_{y}$ as a function of $\kappa$ is that the magnitude of the drift velocity, $|s_{y}|$, has a maximum at some $\kappa = \kappa_{m} > 1$ (since $s_{y}|_{\pi + \phi} = -s_{y}|_{\phi}$, the velocity direction depends on $\phi$). In particular, if $\phi = 0$ and $\alpha = 10$, then $\kappa_{m} \approx 1.25$ and $(s_{y}/ \gamma) |_{\kappa = \kappa_{m}} \approx -0.44$ (see the curve for $\alpha = 10$ in Fig.~\ref{fig7}). If, in addition, $\gamma = 0.1$ and $v_{m} = 10^{-2}\; \mathrm{cm/s}$ \cite{DePe2017}, then from the dimensional drift velocity $v_{\mathrm{dr}} = v_{m}s_{y}$ one finds $v_{\mathrm{dr}} |_{\kappa= \kappa_{m}} \approx -4.4 \times 10^{3}\; \mathrm{nm/s}$. It should also be noted that while $s_{y}$ at $\kappa > \kappa_{m}$ is a monotonic function of $\kappa$, this is not the case for $d s_{y}/d \kappa$ (for illustration, see curves for $\alpha = 10$ and $\alpha = 50$). Finally, using (\ref{s_y1}), it can be verified that $s_{y} \sim (1/\pi^{2}) \gamma \kappa\alpha \sin{\phi}$ as $\kappa \to 0$, i.e., the asymptotic formula (\ref{lim2}) holds for both $\alpha \to 0$ and $\kappa \to 0$, and the asymptotic formula (\ref{lim1}) well describes the dependence of $s_{y}$ on $\kappa$ for $\alpha = 50$ and $\kappa <1$.

\section{Conclusions}
\label{Concl}

We have presented analytical results on the drift of suspended ferromagnetic particles induced by the Magnus force. Our theoretical approach is based on a minimal set of coupled equations for the translational and rotational motions of single-domain particles caused by a harmonic force and a non-uniformly rotating magnetic field. In the approximation of small Reynolds numbers and frozen magnetization, the translational equation of motion represents the instantaneous particle velocity which, due to the Magnus effect, depends on the particle rotation. In contrast, the rotational equation of motion does not depend on the particle velocity and reduces to the equation describing the periodically driven overdamped pendulum. Within this framework, the drift velocity, which is the most important characteristic of the drift motion of particles, is completely determined by the steady-state solution of that equation.

In this paper we have derived an exact steady-state solution of the equation for an overdamped pendulum driven by the square wave (this is our main analytical result). It represents the lag angle between the magnetic field and the particle magnetization and depends on only two dimensionless parameters $\alpha$ and $\kappa$ that are associated with the inverse rotational relaxation time and the square wave amplitude, respectively. We have established that there are two different modes of particle rotation. The first occurs at $\kappa \leq 1$ and is characterized by the lag angles belonging to the interval $(-\pi, \pi)$ for all values of the parameter $\alpha$. In contrast, the lag angles interval in the second mode, which occurs at $\kappa >1$, grows infinitely with increasing $\alpha$. This mode exists when the maximum value of the azimuthal angle of the magnetic field exceeds the critical value $\alpha/4$.

Using the analytical expressions for the lag angle in these modes, we have calculated the drift velocity of particles and analyzed its dependence on the model parameters. One of the most interesting features of the drift velocity is that it varies non-monotonically with $\alpha$. Moreover, if the initial phase of the magnetic field is chosen appropriately, then there always exists a critical value $\alpha_{\mathrm{cr}}$ of this parameter at which the drift velocity equals zero. Since in this case the particles with $\alpha < \alpha_{\mathrm{cr}}$ and $\alpha > \alpha_{\mathrm{cr}}$ drift in opposite directions, the phenomenon of bidirectional drift could be used for separation of different particles (e.g., particles with different magnetization) in suspensions. One more important feature of the drift velocity is that its magnitude as a function of the parameter $\kappa$ exhibits a maximum at some $\kappa >1$ (when the second mode of particle rotation is realized). We have confirmed our theoretical predictions by numerical results obtained from basic equations for the translational and rotational motions of suspended particles.

\section*{Acknowledgments}

This work was supported by the Ministry of Education and Science of Ukraine under Grant No.\ 0116U002622.

\end{document}